\documentstyle[11pt,fleqn]{article}
\begin{document}
\thispagestyle{empty}
\date{}
\title{\bf Spectral Analysis of Microwave Background Perturbations
Induced by Cosmic Strings}
\author{Leandros Perivolaropoulos\thanks{Division of Theoretical Astrophysics,
Harvard-Smithsonian Center for Astrophysics
60 Garden St.
Cambridge, Mass. 02138, USA.}
\thanks{also Visiting Scientist, Department of Physics
Brown University Providence, R.I. 02912, U.S.A.}
}
\maketitle

\begin{abstract}
Using a simple analytic model based on seed superposition we obtain the
spectrum of microwave background perturbations induced by cosmic strings on
all angular scales larger than about 2 armin. We assume standard
recombination in an Einstein de Sitter universe with $h=1/2$ and study the
fluctuation
spectrum along a great circle in the sky.
 Doppler and potential
perturbations on the last scattering surface (Sachs-Wolfe effect) are shown to
dominate over
post-recombination perturbations on scales smaller than about 2 degrees. Using
a filter
function corresponding to the COBE experiment we obtain an effective power
spectrum index $n_{eff}\simeq 1.35$ in good agreement with the recently
announced second year COBE data showing $n_{eff}\simeq 1.5$. The only
free parameter ($G\mu$) of the string model is fixed by normalizing on the
COBE detection leading to $G\mu=1.6$. Other parameters (e.g. the {\it rms}
string velocity) are fixed by comparing with string simulations. Using these
parameter values we compare the {\it rms} temperature fluctuations
$({{\Delta T}\over T})_{\it rms}$ predicted for ongoing experiments
(Tenerife, SP91, MAX, OVRO etc) with detections and with the corresponding
predictions of inflationary models.
\end{abstract}

{\it Subject headings:} cosmic microwave background - cosmic strings
\newpage

\section{\bf Introduction}

The rapid development of both observations and theory has turned the search for
the origin of
cosmic structure into one of the most exciting fields of scientific research.
Succesful
observations performed during the past few years have imposed severe
constraints on theories of
structure formation. In spite of these constraints one important question
remains open: {\it What is
the origin of primordial fluctuations that gave rise to structure in the
universe?}
 Two classes of theories attempting to answer this question have emerged during
the past ten years
and have managed to survive through the observational constraints with only
minor adjustments.

According to the first class, primordial fluctuations are produced by quantum
fluctuations of a
linearly coupled scalar field during a period of inflation (Hawking 1982;
Starobinsky 1982; Guth \&
Pi 1982; Bardeen, Steinhardt \& Turner 1983). These
fluctuations are subsequently expected to become classical and provide the
progenitors of structure
in the universe. Because of the extremely small linear coupling of the scalar
field, needed to
preserve the observed large scale homogeneity, the inflationary perturbations
are expected by the
central limit theorem, to obey Gaussian statistics. This is not the case for
the second class of
theories.

According to the second class of theories (Kibble 1976; Vilenkin 1981; Vilenkin
1985; Turok 1989;
Brandenberger 1992),
primordial perturbations are provided by {\it seeds} of trapped energy density
produced during
symmetry breaking phase transitions in the early universe. Such symmetry
breaking is predicted by
Grand Unified Theories (GUT's) to occur at early times as the universe cools
and expands. The
geometry of the produced seeds, known as {\it topological defects} is
determined by the topology of
the vaccuum manifold of the physically realized GUT. Thus the defects may be
pointlike (monopoles),
linelike (cosmic strings), planar (domain walls) or collapsing pointlike
(textures).

The cosmic string theory (Vilenkin 1981) for structure formation is the oldest
and (together with
textures (Turok 1989)) best studied theory of the topological defect class. By
fixing its single free
parameter $G\mu$ ($\mu$ is the {\it effective} mass per unit length of the
wiggly string and $G$ is
Newtons constant) to a value consistent with microphysical requirements coming
from GUT's, the
theory may automatically account for large scale filaments and sheets
(Vachaspati 1986;
Stebbins {\it et. al.} 1987; Perivolaropoulos, Brandenberger \& Stebbins 1990;
Vachaspati \& Vilenkin 1991; Vollick 1992; Hara \& Miyoshi 1993), galaxy
formation at epochs $z\sim 2-3$ (Brandenberger {\it et. al.} 1987) and galactic
magnetic fields
(Vachaspati 1992b). It
can also provide large scale peculiar velocities (Vachaspati T. 1992a;
Perivolaropoulos \& Vachaspati 1993) and is consistent with the amplitude,
spectral index
(Bouchet, Bennett \& Stebbins 1988; Bennett, Stebbins \& Bouchet 1992;
Perivolaropoulos 1993a;
Hindmarsh 1993) and the statistics (Gott {\it et. al.} 1990; Perivolaropoulos
1993b; Moessner,
Perivolaropoulos \& Brandenberger 1993;  Coulson {\it et. al.} 1993; Luo 1993)
 of the cosmic microwave background (CMB) anisotropies measured by the COBE
collaboration
(Smoot {\it et. al.} 1992; Wright {\it et. al.} 1992) on large angular scales
($\theta\sim
10^\circ$).

The CMB observations provide a valuable direct probe for identifying signatures
of cosmic strings.
There are three main mechanisms by which strings can produce CMB temperature
fluctuations.

The first mechanism has been well studied both analytically (Brandenberger \&
Turok 1986;
Stebbins 1988; Veeraraghavan \& Stebbins 1990; Perivolaropoulos 1993a;
Perivolaropoulos 1993b;
Moessner {\it et. al.} 1993) and
using numerical simulations (Bouchet {\it et.al.} 1988;
Bennett {\it et. al.} 1992) and is known as the {\it
Kaiser-Stebbins effect} (Kaiser \&  Stebbins 1984; Gott 1985). According to
this
effect, moving long strings present between the time of recombination $t_{rec}$
and the present time
$t_0$, produce step-like temperature discontinuities between photons that reach
the observer through
opposite sides of the string. These discontinuities are due to the peculiar
nature of the spacetime
around a long string which even though is {\it locally} flat, {\it globally}
has the geometry of
a cone with deficit angle $8\pi G\mu$. The magnitude of the discontinuity is
proportional to the
deficit angle, to the string velocity $v_s$ and depends on the relative
orientation between the unit
vector along the string ${\hat s}$ and the unit photon wave-vector ${\hat k}$.
It is given by
(Stebbins 1988)
\begin{equation}
{{\delta T}\over T}=\pm 4\pi G\mu v_s \gamma_s {\hat k} \cdot ({\hat v_s}\times
{\hat s})
\end{equation}
where $\gamma_s$ is the relativistic Lorentz factor and the sign changes when
the string is
crossed. The angular scale over which this discontinuity persists is given by
the radius of
curvature of the string $\Psi(t)=\xi \Theta(t)/2$ (the parameter $\xi$ measures
the string curvature
radius as a fraction of the horizon angular scale $\Theta(t)$).

The second mechanism is based on the Doppler effect. Moving long strings
present on or before the
last scattering surface produce, due to their deficit angle, velocity fields
directed towards the
surface swept by the string. These perturbations affect both the plasma and the
dark matter.
The growth of velocity perturbations produced before recombination on the
plasma, is prevented by
pressure and photon drag. However, velocity fields produced on the plasma last
scatterers of
photons by moving long strings {\it at} the time of recombination {\it are}
important because they
produce temperature fluctuations on the scattered photons via the {\it Doppler
effect}. The
magnitude of these fluctuations can be obtained from the velocity perturbations
(Vachaspati \& Vilenkin 1991; Vachaspati 1992a) induced by a long string as
\begin{equation}
{{\delta T}\over T}={\hat k}\cdot {\vec v}=\lambda \pi G\mu v_s \gamma_s {\hat
k}\cdot ({\hat v_s}
\times {\hat s})
\end{equation}
where
\begin{equation}
\lambda=(1+{{(1-{T\over \mu})}\over {2<(v_s \gamma_s)^2 >}})
\end{equation}
and T is the tension of the wiggly long string estimated by simulations to be
$T\simeq 0.7\mu$. The
scale over which these fluctuations persist is the radius of curvature of long
strings at
$t_{rec}$ ($\Psi (t_{rec})$).

The third mechanism is based on potential fluctuations on the last scattering
surface produced by
both loops and long string wakes. Wakes are planar density enhancements induced
by the growing
velocity perturbations of long strings on dark matter (Stebbins {\it et. al.}
1987;
Perivolaropoulos {\it et. al.} 1990).
 These perturbations begin to grow at about
the time of equal matter and radiation $t_{eq}$ (assuming Cold Dark Matter
(CDM)). A wake produced
by a long string at time $t_i$ has angular dimensions $\Psi(t_i)\times v_s
\gamma_s \Theta (t_i)$
and a surface density at time t given by (Perivolaropoulos {\it et. al.} 1990)
\begin{equation}
\sigma (t_i,t)=\lambda 4\pi G\mu v_s \gamma_s t_i \rho(t_i) ({t\over
t_i})^{2/3}
\end{equation}
where $\rho(t_i)$ is the mean dark matter density at $t_i$. The temperature
perturbation due to
the potential $\Phi(x,t_{rec})$ (Sachs \& Wolfe 1967; Stebbins 1993) produced
by the wakes present
on the last scattering surface is
\begin{equation}
{{\delta T}\over T}={1\over 3} \Phi_w(x,t_{rec})={1\over 3}4\pi
G\sigma(t_i,t_{rec}) h(x)
\end{equation}
where $h(x)=x\hspace{1mm} \cos\phi$ is the {\it perpendicular} distance to the
wake as a function
of
the distance $x$ from the wake on the last scattering surface and $\phi$ is the
angle between the
photon wave-vector and the wake surface. As in the previous cases, the {\it
angular} distance from
the wake over which this perturbation persists is approximatelly $\Psi(t_i)$.

Loops present on the last scattering surface with their accreted dark matter
are also expected to
contribute to the potential fluctuations. It will be seen however that their
contribution is
minimal due to their very small size indicated by numerical simulations.
According to simulations
(Bennett \& Bouchet 1988) loops are produced with relativistic velocities
(which quickly erase any
initial correlations) and typical size $R(t)\simeq 10^{-4} H^{-1}(t)$
($H^{-1}(t)$ is the Hubble
scale at time $t$), which for $t=t_{rec}$ corresponds to an angular scale of
about 0.3 arcsec (even
though the largest loops can be about ten times larger (Bouchet 1988)). This
scale is too
small for any observable effects in present experiments. For completeness
however we will include
loops in our analysis in order to explicitly demonstrate that they are
unimportant compared to other
types of perturbations.

The temperature fluctuations induced by a loop of radius $R$, due to the
disturbance of the last
scattering surface potential at a distance $x$ from the center of the loop may
be approximated by
\begin{eqnarray}
{{\delta T}\over T}&=&{1 \over 3}\Phi_l (x,t_{rec})\simeq {{\beta G\mu}\over 3}
({t_{rec}\over
t_i})^{2/3} \hspace{1cm} \vert x\vert \leq R \\
{{\delta T}\over T}&\simeq & 0 \hspace{1cm} \vert x\vert >> R
\end{eqnarray}
where $\beta$ is a parameter determining the length of string in a loop of
radius $R$ (typically
$\beta\leq 10$). The time dependent growth factor takes into account the
accretion of CDM (as in
the case of wakes).

\section{\bf Derivation of Spectrum}

The above three mechanisms have specified four distinct types of seed functions
that need to be
superimposed (in different ways) in order to construct the CMB spectrum of
string induced temperature
fluctuations. The next step is to give an expression of the spectrum in terms
of these seed
functions. For simplicity we will focus on perturbations along a great circle
on the sky. By
isotropy, our results for the correlation function and the spectrum $P(k)$ are
valid for {\it any}
such circle and the extension of our results to {\it patches} will be shown to
be straightforward.

Consider a seed function $f_1^\Psi (\theta)$ of size proportional to $\Psi$,
superimposed at random
positions $\theta_n$ and with variable amplitude $a_n$, $N$ times on a circle
($-\pi\leq \theta
<\pi$). The resulting function is
\begin{equation}
f(\theta)=\sum_{n=1}^{N} a_n f_1^\Psi
(\theta-\theta_n)={1\over{2\pi}}\sum_{n=1}^{N} a_n
\sum_{k= -\infty}^{+\infty}{\tilde f_1^\Psi}(k) e^{ik(\theta-\theta_n)}
\end{equation}
where
\begin{equation}
{\tilde f_1^\Psi}(k)\equiv \int_{-\pi}^{+\pi} d\theta f_1^\Psi (\theta) e^{-i k
\theta}
\end{equation}
is the Fourier transform of $f_1(\theta)$. The Fourier transform of $f(\theta)$
is
\begin{equation}
{\tilde f}(k)={\tilde f_1^\Psi}(k)\sum_{n=1}^{N}a_n e^{i k \theta_n}
\end{equation}
The probability distribution of the phases of this pattern has been studied
previously
(Perivolaropoulos 1993c). Here we focus on the  power spectrum of the
resulting pattern which is
\begin{equation}
P_0 (k)\equiv <\vert {\tilde f}(k)\vert^2>=N \vert{\tilde f_1^\Psi}(k)\vert^2
<\vert a_n\vert^2>
\end{equation}
where $<>$ denotes ensemble average.

In a cosmological setup the role of seeds is played by topological defects
which obey a
`scaling solution' and therefore their size is a fixed fraction of the horizon
at any given time.
As the comoving horizon expands by a factor $\alpha$, the size (total number)
of superimposed seeds
on the great circle will also increase (decrease\footnote{since the total
number of horizons
on the circle decreases}) by the same factor. Therefore, after $Q$ expansion
steps the resulting
spectrum will be
\begin{equation}
P_Q (k)=\sum_{q=0}^Q P_q (k)\equiv \sum_{q=0}^{Q} {N\over {\alpha^q}} \vert
{\tilde f_1}^{\alpha^q
\Psi} (k) \vert^2 <\vert a_n\vert^2>
\end{equation}
This is a general result that can be easily applied to any particular type of
seed function $f_1
(\theta)$ provided the following quantities are specified
\begin{enumerate}
\item
{\it The number of seeds N at the first expansion step}. For a scaling defect
with M seeds per
horizon scale we have $N(t_i)=M\times {{2\pi}\over \Theta(t_i)}$ where
$\Theta(t_i)$ is the angular
size of the horizon at the first expansion step. For $H_0=50 km/(sec\cdot
Mpc)$,  ({\it i.e.} h=1/2)
and $\Omega_0=1$ we have $N(t_{rec})\simeq 200\hspace{1mm} M$ and
$N(t_{eq})\simeq 450\hspace{1mm}
M$ since $\Theta (t_{rec})=1.8^\circ$ and $\Theta(t_{eq})=0.8^\circ$.
\item
{\it The maximum and minimum size of superimposed seeds}\\
 $(\Psi_{min}(t_i),\Psi_{max}(t_i))\equiv
(\xi \Theta(t_i)/2,\xi \Theta(t_f)/2)$. For initial and final times of
superposition $(t_i,t_f)=(t_{rec},t_0)$ (to be used for the Kaiser-Stebbins
effect) we have
\begin{equation}
(\Psi_{min},\Psi_{max})=(0.016 \xi, \pi \xi)
\end{equation}
while for $(t_i,t_f)=(t_{eq},t_{rec})$ (to be used for the scales of potential
perturbations) we have
\begin{equation}
(\Psi_{min},\Psi_{max})=(0.007 \xi, 0.016 \xi)
\end{equation}
The number of expansion steps is obtained simply as
\begin{equation}
Q={{\log ({{\Psi_{max}}\over {\Psi_{min}}})}\over {\log \alpha}}
\end{equation}
\item
{\it The comoving horizon expansion factor $\alpha$}. Taking each expansion
step when the {\it
physical} horizon scales by $\delta$ in size leads to $\alpha=\delta^{1/3}$.
Increasing $\alpha$
tends to increase each term in the sum (12) but this effect is compensated by
the decrease in $Q$
due to (15). Thus for reasonable values of $\delta$ ($1.5\leq \delta \leq 3$)
our results are rather
insensitive to the value of $\alpha$. In what follows we use $\delta=2$ which
implies
$\alpha=1.26$.
\item
{\it The magnitude of $a_n$ and the form $f_1(\theta)$} of the seed functions
given by equations
(1), (2), (5) and (6).
\end{enumerate}
We are now in position to determine the contribution to the power spectrum for
each one of the
three mechanisms by which strings can produce temperature fluctuations on the
CMB.
\begin{enumerate}
\item
{\it Kaiser-Stebbins effect:} It is easy to see from (1) that in this case
\begin{eqnarray}
f_1^\Psi (\theta)& = & -1 \hspace{1cm} 0\leq \theta \leq \Psi \\
f_1^\Psi (\theta)& = & +1 \hspace{1cm} 0\geq \theta \geq -\Psi \\
f_1^\Psi (\theta)& = & 0 \hspace{1cm} otherwise
\end{eqnarray}
while
\begin{equation}
a_n=4\pi G\mu (v_s \gamma_s)_n \cos\phi_n
\end{equation}
where $\cos\phi_n={\hat k}\cdot({\hat v_s}\times {\hat s})_n$. It is now easy
to show that
\begin{equation}
{\tilde f_1}^\Psi (k)={{4\sin^2 (k\Psi/2)}\over k}
\end{equation}
and
\begin{equation}
<\vert a_n \vert^2>={1\over 3} (4\pi G\mu)^2 <(v_s \gamma_s)^2>
\end{equation}
Also, since $(t_i,t_f)=(t_{rec},t_0)$, $(\Psi_{min},\Psi_{max})$ is given by
(13), $Q=23$ (from
(15) with $\alpha = 1.26$) and $N=200M$. Thus, substituting in (12) we obtain
\begin{equation}
P_{KS}(k)=4\times 10^4  \hspace{1mm} (G\mu)^2 \hspace{1mm} b \hspace{1mm}
\sum_{q=0}^{Q=23}
{{\sin(0.008 \hspace{1mm}\xi \hspace{1mm}\alpha^q\hspace{1mm} k)}\over
{\alpha^q k^2}}
\end{equation}
where $b\equiv M<(v_s \gamma_s)^2>$.
\item
{\it Potential perturbations:}
\begin{enumerate}
\item
{\it Wakes:}  We are interested in all the wakes present on the last scattering
surface thus
causing potential perturbations on it. Those wakes were formed at time $t_i$,
with
$t_{eq} \leq t_i \leq t_{rec}$ (since perturbations start to grow at $t_{eq}$).
All these wakes must
be taken into account with their growth factors. In the case of wakes, the sum
of equation
(12) will run not on the epoch that the photon was affected (as was the case
for the Kaiser-Stebbins
component of the spectrum) but on the growth factor and the size of the wake
present on the last
scattering surface. In this case, we have
$(t_i,t_f)=(t_{eq},t_{rec})$ which implies $Q=3.6$ and $N=450M$. Using (5) and
following the same
steps as for the Kaiser-Stebbins effect we obtain
\begin{equation}
P_w (k)=1.3\times 10^2 \hspace{1mm} (G\mu)^2 \hspace{1mm} b \hspace{1mm}
\xi^4\hspace{1mm} \sum_{q=0}^4 {1\over
{\alpha^{3q}}} ({{\sin g} \over g}-2({{\sin g} \over g})^2)^2
\end{equation}
where
\begin{equation}
g=7\times 10^{-3} \hspace{1mm} \xi \hspace{1mm} k \hspace{1mm} \alpha^q
\end{equation}

\item
{\it Loops:} Using (6), (7) and (12) with $(t_i,t_f)=(t_{eq},t_{rec})$ we
obtain
\begin{equation}
P_l (k)= 5.6 \hspace{1mm} \beta^2 \hspace{1mm} (G\mu)^2 \hspace{1mm}
M_l\hspace{1mm}
\sum_{q=0}^4 {{\sin^2 (0.014\hspace{1mm} k\hspace{1mm} \xi_{l}\hspace{1mm}
\alpha^q)}\over
{\alpha^{5q} k^2}}
\end{equation}
where $M_l$ is the cubic root of the number of loops per horizon scale (i.e.
the number of loops
that intersect a horizon size arc on the sky) and $\xi_{l}$ is typical size of
loops as a
fraction of the horizon scale. The exact values of the parameters $\beta$,
$\xi_l$ and $M_l$ are
unimportant for our calculations, since it will be seen that for all values of
parameters which are
consistent with string simulations, the term $P_l (k)$ is orders of magnitude
smaller than the
other terms contributing to the power spectrum for $k\leq 10^4$ ($\theta \geq
1$ arcmin). For
definiteness we will use $\beta=10$, $M_l=40$ and $\xi_l=10^{-3}$ (values which
tend to overestimate
the contribution of loops). Even with these values it will be seen that loops
are unimportant.
\end{enumerate}
\item
{\it Doppler effect:} Since velocity perturbations on the plasma do not grow
before recombination
we will only consider the contribution of strings present at the time of
recombination. Using (2)
and (12) with $t_i=t_f=t_{rec}$ we obtain
\begin{equation}
P_D (k)=1.2\times 10^4 \lambda^2 (G\mu)^2 b {{\sin^2 (0.016 \hspace{1mm} \xi
\hspace{1mm} k)}\over
k^2} \end{equation}
\end{enumerate}
The total spectrum is therefore given by
\begin{equation}
P(k) = P_{KS} (k) + P_w (k) + P_l (k) + P_D (k)
\end{equation}

\section{\bf Parameter Fixing-Scale Invariance}

Before being able to make predictions about ongoing CMB experiments we must
determine the only free
parameter $G\mu$ as well as the parameters $b$, $\lambda$ and $\xi$.
String simulations (Bennett \& Bouchet 1988; Allen \& Shellard 1990) indicate
that
$M\simeq 10$ while $(v_s\gamma_s)_{rms} \simeq 0.15-0.2$ implying $b\simeq
0.24$ and $\lambda\simeq
6$. We will verify these values by directly fitting our spectrum with partial
CMB
spectra obtained by simulations on large angular scales. Bouchet, Bennett and
Stebbins (hereafter
BBS) have used numerical simulations to calculate the term $P_{KS} (k)$ for a
single expansion step.
Their result for the total power on angular scales smaller than $\theta_*$ with
$t_i=t_{rec}$,
$t_f\simeq 2 \hspace{1mm} t_{rec}$ is
\begin{eqnarray} P_{BBS} (\theta \leq
\theta_*,\Theta_i=\Theta_{rec})&=&\int_{2\pi/\theta_*}^\infty
{{d^2k}\over{(2\pi)^2}} P_{BBS} (k)
\nonumber\\  &=&(6G\mu)^2
({{\theta_*^{1.7}}\over{0.0012+\theta_*^{1.7}}})^{0.7}
\end{eqnarray}
Our analysis, focusing on a line across the sky rather than a patch predicts
\begin{equation}
P_{an}(\theta \leq \theta_*,\Theta_i=\Theta_{rec})=2\hspace{1mm}
\int_{2\pi/\theta_*}^\infty
{{dk}\over{(2\pi)}} P_{KS}^{Q=0}(k)
\end{equation}
In Figure 1 we show $P_{an}(\theta_*)$ for $b=0.237$ and $\xi=0.45$ (continous
line) superimposed on
$P_{BBS}(\theta_*)$ (dashed line). We also show the $1\sigma$ errors to
$P_{an}(\theta_*)$ obtained
from the variance of $a_n^2$ (dotted lines). The values $b=0.237$ and
$\xi=0.45$ were chosen in order to obtain
the best fit to $P_{BBS}$ but they are in very good agreement with the expected
values
(obtained for $M\simeq 10$, $(v_s \gamma_s)_{rms}\simeq 0.15$ and string radius
of curvature about
half the horizon scale).

Figure 2a shows a superposition of the components of the spectrum (equations
(22), (23), (25)
and (26)) with the above choice of parameters. The sums were performed using
{\it
Mathematica} (Wolfram 1991). Clearly the Kaiser-Stebbins term (continous line)
dominates on large
angular scales ($\theta > 4^\circ$) while the Doppler term (long dashed line)
is dominant on smaller
scales. The contribution of potential perturbations by wakes (dotted line) is
less important but is
clearly not negligible especially on scales of a few arcmin ($k\simeq 1500$).
Finally, the
contibution of loops (short dashed line) is negligible on all scales larger
than 2-3 arcmin ($k\leq
8000$). Figure 2b shows the product $kP(k)$ fot the total spectrum (equation
(27)) with $1\sigma$
errors denoted by the dotted lines.

One of the most interesting questions that may be addressed using the spectrum
of Figure 2b is
{\it `What is the effective power spectrum index n, predicted by cosmic strings
on COBE angular
scales?'}. Previous studies (Bennett, Stebbins \& Bouchet 1992;
Perivolaropoulos 1993a) have
addressed this question without taking into account the effects of potential
and Doppler
perturbations. The correlation function $C_1(\theta)$ for perturbations along a
great circle is given
in terms of $P(k)$ as
\begin{equation}
C_1 (\theta)=<{{\delta T}\over T}(\phi) {{\delta T}\over
T}(\theta+\phi)>_\phi={1\over{(2\pi)^2}}\sum_{k=-\infty}^{k=+\infty} P(k) e^{i
k \theta}
\end{equation}
For 2d maps the corresponding equation is ($l\gg1$) (Efstathiou 1989)
\begin{equation}
C_2(\theta)\simeq{1\over{(2\pi)^2}}\int d^2l \hspace{1mm} C_l \hspace{0.5mm}
e^{i {\vec l}\cdot {\vec \theta}}
\end{equation}
By isotropy we must have $C_1 (\theta)=C_2 (\theta)$. It may also be shown that
$l^2 C_l\sim
l^{n-1}$ where $n$ is the power spectrum index. Since both $k$ and $l$ are
Fourier conjugate of
$\theta$ we have $k\simeq l$. Also (30) and (31) imply that $P(k)\simeq \pi l
C_l$ and therefore
\begin{equation}
kP(k)\sim k^{n-1}
\end{equation}

Figures 3a and 3b show the best linear fit of the log-log plot $kP(k)$ vs $k$,
for $5\leq k\leq 20$
and $5\leq k\leq 100$ respectively.
The best fits give $n=1.35$ (Figure 3a) and $n=1.48$ (Figure 3b). This result
indicates that cosmic
strings favor values of $n$ somewhat larger than 1 in agreement with recent
indications from the
Tenerife experiment and the second year data of COBE (Bennett {\it et. al.}
1994) which favor
$n\simeq 1.5$. In contast it is much harder for inflationary models to explain
such high values of
$n$ (Lyth \& Liddle 1994; Steinhardt 1993).

There is a simple analytic way to show that in the sector of the power spectrum
where the
Kaiser-Stebbins effect dominates, a scale invariant ($n\simeq 1$) spectrum
should be expected.
For ${\tilde f_1^\Psi}(k)\sim \Psi {\tilde f_1^{\Psi=1}}(k\Psi)$ (as in the
case of the
Kaiser-Stebbins seed functions), (12) may be writen as
\begin{equation}
P(k)=\sum_{q=0}^Q P_q (k)=\sum_{q=0}^Q \alpha^q P_0 (\alpha^q k)\simeq \alpha
P(\alpha k),
\hspace{0.3cm} \Psi_{max}^{-1}\leq k \leq \Psi_{min}^{-1}
\end{equation}
Therefore
\begin{equation}
k P(k)\sim {\rm const}
\end{equation}
which indicates a scale invariant spectrum for the Kaiser-Stebbins term in the
angular scale range
$\theta \geq 2^\circ$ ($k\leq \Psi_{min}^{-1}$). The Kaiser-Stebbins term
plotted in Figure 2a
(continous line) is in agreement with this result (the best fit for this
component of the spectrum
is obtained for $n=1.12$).

\section{\bf Predictions-Conclusion}

It is now straightforward to use the derived power spectrum with the
appropriate filter functions
in order to make predictions about ongoing experiments. The predicted $rms$
temperature
fluctuations ${{\Delta T} \over T}_{rms}$ for an experiment with window
function $W(k)$ is
\begin{equation}
{{\Delta T} \over T}_{rms}=(C(0))^{1/2}=[{1\over {2\pi^2}}\sum_{k=0}^\infty
P(k) W(k)]^{1/2}
\end{equation}
For COBE we have ${{\Delta T} \over T}_{rms}\simeq 1.1\times 10^{-5}$ and
$W(k)\simeq
e^{-k^2/18^2}$. Thus (35) may be used to fix the single free parameter $G\mu$.
The result is
\begin{equation}
G\mu\simeq 1.6
\end{equation}
in agreement with previous analyses (Bennett, Stebbins \& Bouchet 1992;
Perivolaropoulos 1993a)
valid on the COBE angular scales ($\theta \sim 10^\circ$). This result,
combined with the fit to the
BBS simulation, completely fixes the predicted power spectrum $P(k)$ and the
corresponding
correlation function $C(\theta)$. Figure 4 shows this correlation function
smoothed by the COBE
filter function, superimposed with the COBE data (Smoot {\it et. al.} 1992;
Wright {\it et. al.} 1992).

Using $W(k)=e^{-(k-k_0)^2/{\Delta k}^2}$ and fixing $k_0$, $\Delta k$ for some
of the
ongoing experiments (TEN: Watson {\it et. al.} 1992; SP91: Gaier T. {\it
et.al.} 1992;
SP91: Schuster {\it et. al.} 1993; MSAM: Cheng {\it et. al.} 1993; SK: Wollack
{\it et. al.} 1993;
MAX: Meinhold {\it et. al.} 1993; MAX: Gunderson {\it et. al.} 1993;
WD: Tucker {\it et. al.} 1993; OVRO: Readhead {\it et.al.} 1989) we
are in position to  predict the corresponding value of ${{\Delta T} \over
T}_{rms}$ thus testing the
cosmic string model. These predictions with $1\sigma$ errors coming from the
variance of $a_n^2$ are
shown in Table 1. We also show some of the detections and upper limits existing
to date as well as
the predictions of inflationary models for $0.8\leq n \leq 1.0$, $\Lambda=0$
(Bond {\it et. al.}
1994). At this time both inflationary models and cosmic strings appear to be
consistent with
detections at the $1\sigma$ level. However, as the quality of observations
improves, this may very
well change in the near future.

{\bf Table 1}:Detections of ${{\Delta T} \over T}_{rms}\times 10^{6}$ and the
corresponding
predictions of the string and inflationary models ($0.8\leq n \leq 1.0$,
$\Lambda=0$) normalized on
COBE. \vskip 0.1cm
\begin{tabular}{|c|c|c|c|c|c|}\hline
Experiment & $k_0$ & $\Delta k$ & Detection & Strings & Inflation \\ \hline
COBE & 0 & 18 &11 $\pm$ 2&11 $\pm$ 3  & 11 $\pm$ 2 \\ \hline
TEN& 20 & 16 &  $\leq$  17 &13 $\pm$ 3  & 9 $\pm$ 1\\ \hline
SP91& 80 & 70 & 11 $\pm$ 5 &20 $\pm$ 5  & 12 $\pm$ 2 \\ \hline
SK&  85 & 60 & 14 $\pm$ 5 &19$\pm$ 4   & 12 $\pm$ 3 \\ \hline
MAX& 180 & 130 & $\leq$  30 ($\mu Peg$) &21 $\pm$ 5.5   &16 $\pm$ 5   \\ \hline
MAX& 180 & 130 & 49 $\pm$ 8 ($GUM$) &21 $\pm$ 5.5  & 16 $\pm$ 5  \\ \hline
MSAM &  300 & 200 & 16 $\pm$ 4   &19 $\pm$ 4    & 24 $\pm$ 6   \\ \hline
OVRO22 &  600 & 350 & - &13 $\pm$ 4    & 17$\pm$ 7  \\ \hline
WD&  550 & 400 &  $\leq$  12  & 17.5 $\pm$ 4.5    & 7 $\pm$2   \\ \hline
OVRO&  2000 & 1400 &  $\leq$  24 & 13.5 $\pm$ 3.5   & 7 $\pm$ 3  \\ \hline
\end{tabular}
\vskip 0.5cm

In conclusion, we have demonstrated, using a simple analytical method, that the
CMB spectrum
predicted by the cosmic string model can be calculated in a straightforward way
including all the
relevant sources of perturbations. We have also shown that our results are
consistent with numerical
simulations even though their validity extends beyond the resolution of present
simulations.
Finally we showed that the predicted power spectrum index is slightly larger
than 1
($n_{eff}\simeq 1.4$) and that the predicted {rms} temperature fluctuations
${{\Delta T} \over
T}_{rms}$ are consistent with detections to this date on all angular scales
larger than 2-3
arminutes.

Our analysis has assumed standard recombination and values of cosmological
parameters
($\Omega_0=1$, $h=1/2$, CDM, $\Lambda=0$). It is important to extend our
results to less standard
cases including reionization or presence of Hot Dark Matter. Work in this
direction is in progress.

\section {Acknowledgements}
\par
I wish to thank R. Brandenberger and R. Moessner for
interesting discussions and for providing helpful comments after reading the
paper.
I also thank Tanmay Vachaspati for stimulating discussions which initiated this
project.
This work was supported by a CfA Postdoctoral Fellowship.

\section{Figure Captions}
\begin{flushleft}
{\large \bf Figure 1:}The total power on scale less than $\theta_*$ produced by
cosmic strings
during one expansion step starting at $t_{rec}$.
 \vskip .5cm
{\large \bf Figure 2a:} The four components of the power spectrum of CMB
perturbations induced by
cosmic strings
 \vskip .5cm
{\large \bf Figure 2b:} The {\it total} power spectrum of string perturbations
along a great
circle on the sky.
   \vskip .5cm
{\large \bf Figure 3a:} The best linear fit to the total power spectrum for $5
\leq k \leq 20$.
\vskip .5cm
{\large \bf Figure 3b:} The best linear fit to the total power spectrum for $5
\leq k \leq 100$.
\vskip .5cm
{\large \bf Figure 4:} The cosmic string predicted correlation function
smoothed on COBE scales. Superimposed are the first year COBE data.
\end{flushleft}
\newpage
\centerline{\bf References}
\vspace{0.5cm}
\begin{flushleft}
Allen B. \& Shellard E. P. S. 1990, Phys.Rev.Lett. {\bf 64}, 119.\\
Albrecht A. \& Stebbins A. 1993. Phys. Rev. Lett. {\bf 69}, 2615.\\
Bardeen J., Steinhardt P. \& M.Turner M. 1983, Phys.Rev. {\bf D28}, 679.\\
Bennett C. L. {\it et. al.} 1994, {\it Cosmic Temperature Fluctuations from
Two\\....Years of COBE
DMR Observations}, submitted to Ap. J. \\....(ASTROPH-9401012).\\
Bennett D. \& Bouchet F. 1988, Phys.Rev.Lett. {\bf 60}, 257.\\
Bennett D., Stebbins A. \& Bouchet F. 1992, Ap.J.(Lett.) {\bf 399}, L5.\\
Bond R., Crittenden R., Davis R., Efstathiou G. \& Steinhardt P.
1994,\\....Phys. Rev. Lett.
{\bf 72}, 13.\\
Bouchet F. R., Bennett D. P. \& Stebbins A. 1988,  Nature {\bf335}, 410.\\
Bouchet F. R. 1988, in `The Formation and Evolution of Cosmic
Strings',\\....ed. by Gibbons G.,
Hawking S. \& Vachaspati T. (Cambridge Univ.\\....Press), p 359.\\
Brandenberger R. 1992, 'Topological Defect Models of Structure
              Formation \\....After the COBE Discovery of CMB Anisotropies',
Brown
              preprint \\....BROWN-HET-881 (1992), publ. in proc. of the
              International School of \\....Astrophysics "D.Chalonge", 6-13
Sept.1992,
              Erice, Italy, ed. \\....N.Sanchez (World Scientific, Singapore,
1993).\\
Brandenberger R., Kaiser N., Shellard E. P. S., Turok N. 1987. Phys.Rev.
\\....{\bf D36}, 335.\\
Brandenberger R. \& Turok N. 1986, Phys. Rev. {\bf D33}, 2182.\\
Cheng E. S. {\it et. al.} 1993,{\it A Measurement of the Medium Scale
Anisotropy \\....in the CMB},
preprint MSAM-93A.\\
Coulson D., Ferreira P., Graham P. \& Turok N. 1993, {\it $\Pi$ in the Sky? CMB
\\....Anisotrtopies from Cosmic Defects}, PUP-TH-93-1429,
\\....HEP-PH/9310322.\\
Efstathiou G. 1989, in 'Physics of the Early Universe', SUSSP 36, 1989,
              ed. \\....J.Peacock, A.Heavens \& A.Davies (IOP Publ., Bristol,
1990).\\
Gaier T. {\it et.al.} 1992, (SP91), Ap. J. Lett. {\bf 398}, L1.\\
Gott R. 1985, Ap. J. {\bf 288}, 422.\\
Gott J. {\it et. al.} 1990, Ap.J. {\bf 352}, 1.\\
Guth A. \& Pi S. -Y. 1982, Phys.Rev.Lett. {\bf 49}, 110.\\
Gunderson J. {\it et. al.} 1993, (MAX), Ap. J. Lett. {\bf 413}, L1.\\
Hara T. \& Miyoshi S. 1993, Ap. J. {\bf 405}, 419.\\
Hawking S. 1982, Phys.Lett. {\bf 115B}, 295.\\
Hindmarsh M. 1993, {\it Small Scale CMB Fluctuations from Cosmic
Strings},\\....DAMTP-93-17,
ASTRO-PH 9307040.\\
Kaiser N. \&  Stebbins A. 1984, {\it Nature} {\bf 310}, 391.\\
Kibble T. W. B. 1976, J.Phys. {\bf A9}, 1387.\\
Luo X. 1993, {\it The Angular Bispectrum of the CMB},
CFPA-93-TH-42,\\....ASTRO-PH 9312004.\\
Lyth D. \& Liddle A. 1994,{\it Observational Constraints on the Spectral
Index},\\....Contribution to
the 1993 Capri CMB Workshop. SUSSEX-AST \\....93/12-1, ASTRO-PH/9401014\\
Meinhold P. {\it et. al.} 1993, (MAX), Ap. J. Lett. {\bf 409}, L1.\\
Moessner R., Perivolaropoulos L. \& Brandenberger R. 1993, {\it A Cosmic
\\....String Specific
Signature on the CMB}, Ap. J. in press,\\....ASTRO-PH/9310001.\\
Perivolaropoulos L. 1993a, Phys.Lett. {\bf B298}, 305.\\
Perivolaropoulos L. 1993b, Phys. Rev. {\bf D48}, 1530.\\
Perivolaropoulos L. 1993c, {\it The Fourier Space Statistics of Seed-like
\\....Cosmological
Perturbations}, M.N.R.A.S. in press, CfA-3591, \\....ASTRO-PH/9309023.\\
Perivolaropoulos L. \& Vachaspati T. 1993,{\it Peculiar Velocities and
\\....Microwave Background
Anisotropies from Cosmic Strings}, Ap. J. Lett. \\....in press, CfA-3590,
ASTRO-PH/9303242.\\
Perivolaropoulos L., Brandenberger R. \& Stebbins A. 1990, Phys.Rev.
              \\....{\bf D41}, 1764.\\
Readhead A.C.S. {\it et.al.} 1989, (OVRO), Ap. J. {\bf 346}, 556.\\
Sachs R. \& Wolfe A. 1967, Ap. J. {\bf 147}, 73.\\
Schuster J. {\it et. al.} 1993, (SP91), Ap. J. Lett. {\bf 412}, L47.\\
Smoot G. {\it et. al.} 1992, (COBE), {\it Ap. J. Lett.} {\bf 396}, L1.\\
Starobinsky A. 1982, Phys.Lett. {\bf 117B}, 175.\\
Stebbins A. {\it et. al.} 1987, Ap. J. {\bf 322}, 1.\\
Stebbins A. 1988, Ap.J. {\bf 327}, 584.\\
Stebbins A. 1993, Ann. N.Y. Acad. Sci. {\bf 688} (Texas/PASCOS Proceedings),
824.\\
Steinhardt P. 1993, private communication.\\
Traschen J., Turok N. \& Brandenberger R. 1986,  Phys. Rev. {\bf D34},
919.\\
Tucker G. S. {\it et. al.} 1993, (WD), Princeton preprint.\\
Turok N. and Brandenberger R. 1986,
 Phys. Rev. {\bf D33}, 2175.\\
Vachaspati T. 1986, Phys. Rev. Lett. {\bf 57}, 1655.\\
Vachaspati T. 1992a, Phys.Lett. {\bf B282}, 305.\\
Vachaspati T. 1992b, Phys. Rev. D{\bf 45}, 3487.\\
Vachaspati T. \& Vilenkin A. 1991. Phys. Rev. Lett. {\bf 67},
1057-1061.\\
Veeraraghavan S. \& Stebbins A. 1990, Ap.J. {\bf
365}, 37.\\
Vilenkin A. 1981, Phys.Rev. {\bf D23}, 852.\\
Vilenkin A. 1985, Phys.Rep. {\bf 121}, 263.\\
Vollick D. N. 1992, Phys. Rev. D{\bf 45}, 1884.\\
Watson R. A. {\it et. al.} 1992, (TEN), Nature {\bf 357}, 660.\\
Wright E. L. {\it et. al.} 1992, {\it Ap. J. Lett.} {\bf 396}, L5.\\
Wolfram S. 1991, {\it Mathematica version 2.0},
Addison-Wesley.\\
Wollack E. J. {\it et. al.} 1993, (SK), Ap. J. Lett. {\bf 419}, L49.
\end{flushleft}
\end{document}